\documentclass[submission,copyright,creativecommons]{eptcs}
\providecommand{\event}{LSFA 2011} 
\usepackage{breakurl}             
\usepackage{amsmath,amssymb,amsfonts}
\usepackage[all]{xy}
\usepackage{xcolor}
\usepackage{pdfsync}
\usepackage[normalem]{ulem}

\newtheorem{definition}{Definition}
\newtheorem{theorem}{Theorem}
\newtheorem{lemma}[theorem]{Lemma}

\title{A Formalization of the Theorem of Existence of First-Order Most
  General Unifiers\thanks{Work supported by CNPq Universal Grant 481783/2010-5 and FAPDF PRONEX Grant 2009/00091-0.}}
\author{Andr\'eia B Avelar$^1$\thanks{Author supported by the Brazilian
    research agency CNPq.}, Andr\'{e} L Galdino$^3$, Fl\'avio LC de
  Moura$^{2}$\thanks{Author currently taking one-year leave at Laboratoire PPS,  Universit\'e Paris-Diderot, France.}~
   and Mauricio Ayala-Rinc\'{o}n$^{1,2}$\thanks{Corresponding author, partially supported
    by the Brazilian research agency CNPq.}
\institute{Departamentos de $^1$Matem\'atica e $^2$Ci\^encia da Computa\c{c}\~ao,
  Universidade de Bras\'ilia, Bras\'ilia, Brazil \\
$^3$Departamento de Matem\'atica, 
  Universidade Federal de Goi\'{a}s, Catal\~ao, Brazil} 
\email{\{andreia@mat., flaviomoura@, galdino@, ayala@\}unb.br}
}

\def\titlerunning{A Formalization of the Theorem of Existence of First-Order m.g.u.'s}
\def\authorrunning{A.B. Avelar, A.L. Galdino, F.L.C. de Moura and M. Ayala-Rinc\'on}
\def\copyrightholders{Avelar, Galdino, de Moura and Ayala-Rinc\'on}
\begin{document}
\maketitle

\begin{abstract}
  This work presents a formalization of the theorem of existence of
  most general unifiers in first-order signatures in the higher-order
  proof assistant PVS. The distinguishing feature of this
  formalization is that it remains close to the textbook proofs that
  are based on proving the correctness of the well-known Robinson's
  first-order unification algorithm. The formalization was applied
  inside a PVS development for term rewriting systems that provides a
  complete formalization of the Knuth-Bendix Critical Pair theorem,
  among other relevant theorems of the theory of rewriting. In
  addition, the formalization methodology has been proved of practical
  use in order to verify the correctness of unification algorithms in
  the style of the original Robinson's unification algorithm.
\end{abstract}

\section{Introduction}

A formalization in the proof assistant PVS of the theorem of existence
of most general unifiers (mgu's) in first-order theories is
presented. There are several applications of this theorem on
computational logic, which range from the correctness of first-order
resolution \cite{Ro65}, the correctness of the Knuth-Bendix completion
algorithm \cite{KnBe70} to the correctness of principal type
algorithms \cite{Hi69} and their implementations in programming and
specification languages. This well-known result is stated as follows:

\begin{theorem}[Existence of mgu's]
  \label{existence-mgu}
  Let $s$ and $t$ be terms. Then, if $s$ and $t$ are unifiable then
  there exists an mgu of $s$ and $t$.
\end{theorem}

The analytic proof of this theorem is constructive and the first proof
was introduced by Robinson himself in \cite{Ro65}. In Robinson's
seminal paper, the unification algorithm either gives as output a most
general unifier for each unifiable pair of terms, or fails when there
are no unifiers.  Essentially, the proof of correctness of this
algorithm consists in, firstly, proving that the algorithm always
terminates and, secondly, proving that, when it terminates and returns
an mgu it implies the existence theorem.

Several variants of this first-order unification algorithm appear in
well-known textbooks on computational and mathematical logic,
semantics of programming languages, rewriting theory, type theory
etc. (e.g., \cite{Llo87,EbFlTh84,Bu98,Te2003,BaNi98, Hi97}). Since the
presented formalization follows the classical proof schema, only a
sketch of this proof will be given here.

The development of the PVS \emph{theory} {\tt \bf unification} was
motivated by the formalization of a PVS library for term rewriting
systems \cite{GaAR2008b} in which the theorem of existence of
mgu's is essential in order to obtain complete
formalizations of relevant results such as the well-known
Knuth-Bendix(-Huet) Critical Pair theorem \cite{GaAR2010}.  In
addition to this application of the formalization of the theorem of
existence of mgu's, in \cite{AMARG2010} it was reported a general
verification methodology of first-order unification algorithms,
illustrated through the formalization of the correctness of a greedy
version of Robinson's unification algorithm, that follows the lines of
the formalization of the theorem of existence of mgu's presented in
this paper, in order to check \emph{termination} and \emph{soundness}
of the algorithm. Essentially, in that work it is illustrated how the
verification of \emph{completeness} of a unification algorithm depends
on the particular way in that the algorithm deals with the detection
of non unifiable inputs. But also, in the exercise of formalization of
correctness of efficient unifications algorithms, it is of main
relevance the specific data types and refined strategies used to
efficiently detect and solve differences appearing among the terms
being unified.

In Sec. \ref{Sec:specification}, the necessary analytic concepts
(terms, subterms, positions and substitutions) together with their
corresponding specifications in PVS are given. The formalization of
the theorem of existence of mgu's is presented in
Sec. \ref{Sec:formalization}. Also in Sec.  \ref{Sec:formalization} it
is illustrated how specific unification algorithms {\em \`a la}
Robinson are verified using this methodology. In the sequel related
work and conclusions are presented. The PVS files of the formalization
of the theorem of existence of mgu's and verification of Robinson's
style unification algorithms are available as part of the theory for
term rewriting systems ({\tt trs}) in the NASA LaRC PVS libraries
{\small\tt
  http://shemesh.larc.nasa.gov/fm/ftp/larc/PVS-library/pvslib.html}.

\section{Specification of terms, positions, subterms and
  substitutions}\label{Sec:specification}

Although it is supposed familiarity with unification and its standard
notations (e.g. as in \cite{BaNi98,Te2003}), analytical concepts will
be presented together with their associated specifications in PVS.

Consider a signature $\Sigma$ in which function symbols and their
associated arities are given as well as an enumerable set $V$ of
variables.

\begin{definition}[Well-formed terms]
  The set of well-formed terms, denoted by $T(\Sigma,V)$, over the
  signature $\Sigma$ and the set $V$ of variables is recursively
  defined as: $i)\;\; x\in V$ is a well-formed term and $ii)$ for each
  $n$-ary function symbol $f\in\Sigma$ and well-formed terms
  $t_1,\ldots,t_n$, $f(t_1,\ldots,t_n)$ is a well-formed term.
\end{definition}

Note that constants are $0$-ary well-formed terms.

In the sequel, for brevity ``terms'' instead of ``well-formed terms''
will be used.

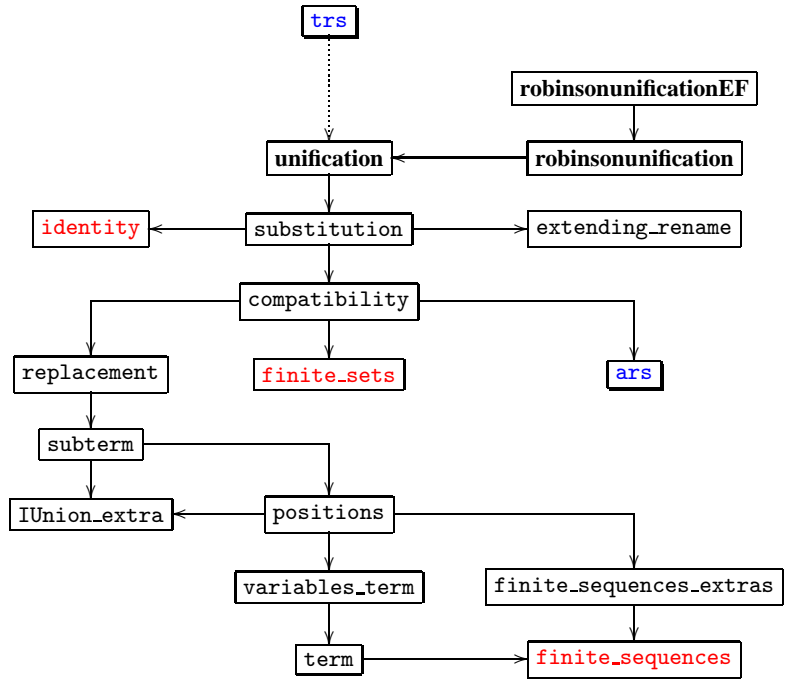
\begin{figure*}[ht!]
  \makebox[\textwidth][c]{ \xymatrix@R14pt{ &
      *+[F-,]{\txt{{\footnotesize{\color{blue}\tt trs}}}}\ar@{.>}[dd]
      &
      \\
      & & *+[F]{\txt{{\footnotesize{\tt \bf robinsonunificationEF}}}}\ar[d]\\
      & *+[F]{\txt{{\footnotesize{\tt \bf unification}}}}\ar[d] &
      *+[F]{\txt{{\footnotesize{\tt \bf robinsonunification}}}}\ar[l]
      \\
      *+[F]{\txt{{\footnotesize{\tt \color{red}{identity}}}}} &
      *+[F]{\txt{{\footnotesize{\tt substitution}}}}\ar[d]\ar[l]\ar[r]
      & *+[F]{\txt{{\footnotesize{\tt extending\_rename}}}}
      \\
      *{}\ar[d]& *+[F]{\txt{{\footnotesize{\tt
              compatibility}}}}*{}\ar[d]\ar@{-}[l]\ar@{-}[r]
      &*{}\ar[d] &
      \\
      *+[F]{\txt{{\footnotesize{\tt replacement}}}}*{}\ar[d] &
      *+[F]{\txt{{\footnotesize{\tt \color{red}{finite\_sets}}}}} &
      *+[F-,]{\txt{{\footnotesize{\tt \color{blue}{ars}}}}} &
      \\
      *+[F]{\txt{{\footnotesize{\tt subterm}}}}*{}\ar[d]\ar@{-}[r]
      &*{}\ar[d] &
      \\
      *+[F]{\txt{{\footnotesize{\tt IUnion\_extra}}}} &
      *+[F]{\txt{{\footnotesize{\tt
              positions}}}}*{}\ar[d]\ar[l]\ar@{-}[r] & *{}\ar[d]
      \\
      & *+[F]{\txt{{\footnotesize{\tt variables\_term}}}}*{}\ar[d] &
      *+[F]{\txt{{\footnotesize{\tt
              finite\_sequences\_extras}}}}*{}\ar[d]
      \\
      & *+[F]{\txt{{\footnotesize{\tt term}}}}*{}\ar[r] &
      *+[F]{\txt{{\footnotesize{\tt \color{red}{finite\_sequences}}}}}
    }}
  \caption{Hierarchy of {\tt unification} inside the \emph{theory}
    {\tt \color{blue} trs}}
  \label{figUnificationInsideTRS}
\end{figure*}

The hierarchy of the \emph{theory} {\tt\bf unification} is presented
in Fig. \ref{figUnificationInsideTRS}. This is part of the
\emph{theory} {\color{blue}\tt trs} for term rewriting systems
presented in \cite{GaAR2008b}, which includes also the \emph{subtheory} {\tt\color{blue} ars} for abstract reduction systems
\cite{GaAR2008c}.
The most relevant notions related with unification are inside the
\emph{subtheories} {\tt positions, subterm} and {\tt substitution}.
The PVS notions used for specifying these basic concepts are taken
from the prelude \emph{theories} for {\tt finite\_sequences} and {\tt
  finite\_sets}. Finite sequences are used to specify
well-formed terms which are built from variables and function symbols
with their associated arities. This is done by application of the PVS
{\tt DATATYPE} mechanism which is used to define recursive types.

{\small
\begin{verbatim}
term[variable: TYPE+, symbol: TYPE+, arity: [symbol -> nat]] : DATATYPE
BEGIN  vars(v:variable): vars?
       app(f:symbol, args:{args:finite_sequence[term] | args`length=arity(f)}): app?
END term
\end{verbatim}}

  Notice that the fact that a term is well-formed, that is, that
  function symbols are applied to the right number of arguments is
  guaranteed by typing the arguments of each function symbol {\tt f}
  as a finite sequence of length {\tt arity(f)}.

  Finite sets and sequences are also used to specify sets of subterms
  and sets of term positions, as is shown below.

  \subsection{The \emph{subtheories} {\tt positions} and {\tt
      subterm}}

  As usual, positions of a term are defined as finite sequences of
  positive naturals, which simplifies the definitions of subterms and
  occurrences.  A dot ``$\cdot$'' is used for the operation of
  concatenation of two naturals $m$ and $n$, $m\cdot n$, and for the
  concatenation of the elements in sets of naturals; that is $N\cdot M
  :=\{n\cdot m \; |\; n\in N, m\in M\}$. For simplicity $n\cdot M$
  denotes $\{n\}\cdot M$.

\begin{definition}[Positions, subterms, occurrences]
  The set of positions of a term $t$ in $T(\Sigma,V)$, denoted as
  $Pos(t)$, is defined inductively as $i)\;\; Pos(x) :=
  \{\varepsilon\}$ and $ii)\;\; Pos(f(t_1,\ldots, t_n)) :=
  \{\varepsilon\}\;\;\cup\;\; \bigcup_{i=1}^{n} i\cdot Pos(t_i)$,
  where $\varepsilon$ denotes the empty sequence that represents the
  \emph{root position} of the term $t$.
 The subterm at a given position $\pi\in Pos(t)$ of a term $t$ is
  defined inductively as $i)\;\; t|_\varepsilon := t$ and $ii)\;\;
  f(t_1,\ldots, t_n)|_{i\cdot \pi} := t_i|_\pi$.

  The set of subterms of a term $t$ is the set $\{t|_\pi \; |\; \pi\in
  Pos(t)\}$.

  Whenever $s=t|_\pi$, it is said that there is an occurrence of the
  subterm $s$ of $t$ at position $\pi$. The set of positions of
  occurrences of a term $s$ in $t$ is given by the set $\{ \pi \; |\;
  t|_\pi = s\}$.
\end{definition}

The (finite) set of positions {\tt positionsOF} of a term {\tt t} is recursively
specified on its structure as below, where {\tt only\_empty\_seq} is
a set containing an empty finite sequence only, that is the set
containing the root position only.

{\small
\begin{verbatim}
positionsOF(t: term): RECURSIVE positions =
  (CASES t OF  vars(t): only_empty_seq,
            app(f, st): IF length(st) = 0 THEN only_empty_seq 
                        ELSE union(only_empty_seq,
                              IUnion((LAMBDA (i: upto?(length(st))):
                                catenate(i, positionsOF(st(i-1)) ))))
                        ENDIF   ENDCASES)
MEASURE t BY << 
\end{verbatim}}

  \noindent where the operator {\tt IUnion} builds the union of all
  sets of positions of the arguments of a functional term {\tt app(f,
    st)} in which {\tt f} is the name of the function and {\tt st} is
  the sequence of arguments, that is a sequence of length equal to the
  arity of {\tt f}. The positions of the {\tt i}$^{th}$ argument are
  prefixed by {\tt i} in order to build the sequence of positions
  inside this argument relative to the whole term.
 
  Several necessary results on terms, subterms and positions are
  formalized by induction on the structure of terms following the
  lines of this abstract datatype specification. For instance,
  properties, such as the one that states that the set of positions of
  a term is finite as well as the
  one that states that the set of variables occurring in a term is
   finite (lemma {\tt vars\_of\_term\_finite} in the
   \emph{subtheory} {\tt subterm}), and that terms with the same heading symbol
  (applications) have the same number of arguments, presented below,
  are proved by structural induction on the abstract datatype for
  terms.

  {\small
\begin{verbatim}
positions_of_terms_finite : LEMMA is_finite(positionsOF(t))

equal_symbol_equal_length_arg : LEMMA
 FORALL (s, t: term, fs, ft: symbol, 
         ss:{args: finite_sequence[term] | args`length = arity(fs)},
         st:{argt: finite_sequence[term] | argt`length = arity(ft)}) :
   (s = app(fs, ss) AND t = app(ft,st) AND fs = ft) => ss`length = st`length
\end{verbatim}
  }

  For ${\tt p} \in Pos({\tt t})$, in the \emph{subtheory} {\tt
    subterm}, the subterm of {\tt t} at position {\tt p} also is
  specified in a recursive way (now on the length of {\tt p}), as
  follows:

  {\small
\begin{verbatim}
subtermOF(t: term, (p: positions?(t))): RECURSIVE term =
 IF length(p) = 0 THEN t ELSE LET st = args(t), i = first(p), q = rest(p) IN
     subtermOF(st(i-1), q) ENDIF
MEASURE length(p)
\end{verbatim}}

    \noindent where {\tt first} and {\tt rest} are constructors that
    return, respectively, the first element and the rest of a finite
    sequence, and {\tt positions?(t)} is the (dependent) type of all
    positions in {\tt t}, which is specified as follows:

    {\small
\begin{verbatim}
positions?(t: term): TYPE = {p: position | positionsOF(t)(p)}
\end{verbatim}}

      Other results are formalized
      by induction on the length of (sequences representing)
      positions; for instance the ones below stating the equality
      $t|_{p.q} = (t|_p)|_q$ and that whenever $p$ is a position of
      $t$ and $q$ a position of $t|_p$, $p.q$ is a position of $t$,
      are proved by structural induction on terms.

      {\small
\begin{verbatim}
pos_subterm: LEMMA  FORALL (p, q: position, t: term):
     positionsOF(t)(p o q) => subtermOF(t, p o q) = subtermOF(subtermOF(t, p), q)

pos_o_term: LEMMA  FORALL (p, q: position, t: term):
     positionsOF(t)(p) & positionsOF(subtermOF(t, p))(q) => positionsOF(t)(p o q)
\end{verbatim}}

        \subsection{The \emph{subtheory} {\tt substitution} }

        By using the definition of position, the notion of replacement
        of a subterm of a term is stated easily.

\begin{definition}[Replacement of subterms]
  Consider $t\in T(\Sigma,V)$ and $\pi \in Pos(t)$. The term resulting
  from replacing the subterm at position $\pi$ of $t$ by the term $s$
  is denoted by $t[\pi \leftarrow s]$.
\end{definition}

Alternatively, the notation $t[s]_\pi$ is also frequently used in the
literature.

\begin{definition}[Substitution]
  A substitution $\sigma$ is defined as a function from $V$ to
  $T(\Sigma, V)$, such that the domain of $\sigma$, defined as the set
  of variables $\{x \;|\; x\in V, x\sigma\not= x\}$ and denoted by
  $Dom(\sigma)$, is finite.
\end{definition}

\begin{definition}[Homomorphic extension of a substitution]
  The homomorphic extension of a substitution $\sigma$, denoted as
  $\hat{\sigma}$, is inductively defined over the set $T(\Sigma, V)$
  as $i)\;\; x\hat{\sigma}:= x\sigma$ and $ii)\;\;
  f(t_1,\ldots,t_n)\hat{\sigma}:=
  f(t_1\hat{\sigma},\ldots,t_n\hat{\sigma})$.
\end{definition}

Given the notion of homomorphic extension, it is possible to define
substitution composition.

\begin{definition}[Composition of substitutions]
  Consider two substitutions $\sigma$ and $\tau$, their composition
  $\sigma\circ \tau$ is defined as the substitution $\sigma\circ\tau$
  such that $Dom(\sigma\circ\tau) = Dom(\sigma) \cup Dom(\tau)$ and
  for each variable $x$ in this domain, $x(\sigma\circ\tau) :=
  (x\tau)\hat{\sigma}$.
\end{definition}

The \emph{subtheory} {\tt substitution} specifies the algebra of
substitutions. In this \emph{subtheory} the type of substitutions is
built as functions from variables to terms {\tt sig : [V -> term]},
whose domain is finite: {\tt Sub?(sig): bool = is\_finite(Dom(sig))}
and {\tt Sub: TYPE = (Sub?)}. Also, the notions of domain, range, and
the variable range are specified, closer to the usual theory of
substitution as presented in well-known textbooks (e.g.,
\cite{BaNi98}).  These notions are specified as follows:

{\small
\begin{verbatim}
Dom(sig): set[(V)] = {x: (V) | sig(x) /= x}
\end{verbatim}
} {\small
\begin{verbatim}
Ran(sig): set[term] = {y: term | EXISTS (x: (V)): member(x, Dom(sig)) &  y = sig(x)}
\end{verbatim}
} {\small
\begin{verbatim}
VRan(sig): set[(V)] = IUnion(LAMBDA (x | Dom(sig)(x)): Vars(sig(x)))
\end{verbatim}
}

\noindent where {\tt (V)} denotes the type of all terms that are
variables and {\tt Vars(t)} denotes the set of all variables occurring
in a term {\tt t}.

Also, in the \emph{subtheory} {\tt substitution} the homomorphic
extension {\tt ext(sig)} of a substitution {\tt sig} is specified
inductively over the structure of terms:

{\small
\begin{verbatim}
 ext(sigma)(t): RECURSIVE term =   
 CASES t OF 
   vars(t): sigma(t),
   app(f, st):    IF length(st) = 0 THEN t ELSE LET sst = (# length := st`length,
       seq := (LAMBDA (n: below[st`length]): ext(sigma)(st(n)))#) IN app(f, sst) ENDIF
  ENDCASES
 MEASURE t BY <<
\end{verbatim}
}

The composition of two substitutions, denoted by {\tt comp}, is
specified as

{\small
\begin{verbatim}
comp(sigma, tau)(x: (V)): term = ext(sigma)(tau(x))
\end{verbatim}
}

In standard rewriting notation, the homomorphic extension of a
substitution $\sigma$ from its domain of variables to the domain of
terms is denoted by $\hat{\sigma}$, but to simplify notation, usually
textbooks do not distinguish between a substitution $\sigma$ and its
extension $\hat{\sigma}$.  In the formalization this distinction
should be maintained carefully.  For instance observe the following
lemma and its formalization.
 
\begin{lemma}
  Let $s$ be term, $p$ a position of $s$ and $\sigma$ a
  substitution. Then $(s\hat{\sigma})|_{p} = (s|_{p})\hat{\sigma}$.
\end{lemma}

{\small
\begin{verbatim}
subterm_ext_commute: LEMMA FORALL (p: position, s: term, sigma: Sub):
      positionsOF(s)(p) =>  subtermOF(ext(sigma)(s), p) = ext(sigma)(subtermOF(s, p))
\end{verbatim}
}

Several important results useful for the development of
\emph{subtheory} {\tt \bf unification} were formalized in the
\emph{subtheory} {\tt substitution}, e.g., the property that states
that the application of a homomorphic extension of a substitution
preserves the original set of positions of the instantiated term,
formalized as:

{\small
\begin{verbatim}
ext_preserv_pos: LEMMA FORALL (p: position, s: term, sigma: Sub):
           positionsOF(s)(p) => positionsOF(ext(sigma)(s))(p)
\end{verbatim}
}

The lemma below formalizes the set of positions of the instantiation
of a term by a substitution.

{\small
\begin{verbatim}
 positions_of_ext: LEMMA positionsOF(ext(sigma)(t)) = 
   union({p | positionsOF(t)(p) & (NOT vars?(subtermOF(t, p)))}, 
         {q | EXISTS p1, p2: q = p1 o p2 AND positionsOF(t)(p1) AND 
        vars?(subtermOF(t, p1)) AND positionsOF(ext(sigma)(subtermOF(t, p1)))(p2)})
\end{verbatim}}

  Additional formalized lemmas, presented below, state that all
  variables in the domain but not in the range of a substitution
  $\sigma$ disapear in all $\sigma$ instantiated terms and that
  non-variable subterms, i.e. function symbols, remain untouched after
  any possible instantiation.

  {\small
\begin{verbatim}
vars_subst_not_in: LEMMA FORALL t, sigma, x: 
  Dom(sigma)(x) AND (FORALL r: Ran(sigma)(r) => NOT member(x, Vars(r)))
   => NOT member(x, Vars(ext(sigma)(t)))
\end{verbatim}
  } {\small
\begin{verbatim}
ext_preserve_symbol : LEMMA FORALL(s:term, sig:Sub, p:position | positionsOF(s)(p)):
       app?(subtermOF(s, p)) => f(subtermOF(s, p)) = f(subtermOF(ext(sig)(s), p)) 
\end{verbatim}
  }
  
  \section{Formalization of first-order
    unification}\label{Sec:formalization}

  The formalization of the existence of first-order mgu's is presented
  and then it is explained how the formalization technology was
  applied to verify a specific unification algorithm. Again,
  definitions and their corresponding specifications are included.
  The \emph{theory} {\tt \bf unification} consists of 57 lemmas from
  which 30 are \emph{type proof obligations} (TCCs) that are lemmas
  automatically generated by the prover during the type checking. The
  specification file has 273 lines and its size is 9.8 KB and of the
  proof file has 11540 lines and 657 KB.

  Two terms $s$ and $t$ are said to be unifiable whenever there exists
  a substitution $\sigma$ such that $s\hat{\sigma} = t\hat{\sigma}$.
  
\begin{definition}[Unifiers]
  The set of unifiers of two terms $s$ and $t$ is defined as $U(s,t)
  := \{ \sigma \; |\; s\hat{\sigma} = t\hat{\sigma}\}$.
\end{definition}

\begin{definition}[More general substitutions]
  Given two substitutions $\sigma$ and $\tau$, $\sigma$ is said to be
  more general than $\tau$ whenever, there exists a substitution
  $\gamma$ such that $\gamma\circ\sigma = \tau$.  This is denoted as
  $\sigma \leq \tau$.
\end{definition}

\begin{definition}[Most General Unifier]
  Given two terms $s$ and $t$ such that $U(s,t)\not=\emptyset$. A
  substitution $\sigma$ such that for each $\tau\in U(s,t)$, $\sigma
  \leq \tau$, is said to be a most general unifier of $s$ and $t$. For
  short it is said that $\sigma$ is an mgu of $s$ and $t$.
\end{definition}

Now, it is possible to state the theorem of existence of mgu's.
 
\begin{theorem}[Existence of mgu's] 
  Let $s$ and $t$ be terms in $T(\Sigma,V)$ built over a signature $\Sigma$. Then,
  $U(s,t)\not=\emptyset$ implies that there exists an mgu of $s$ and
  $t$.
\end{theorem}

The analytic proof of this theorem is constructive and the first
introduced proof was presented by Robinson himself in \cite{Ro65}. In
Robinson's paper, a unification algorithm was introduced, which either
gives as output a most general unifier for each unifiable pair of
terms or fails when there are no unifiers.  The proof of correctness
of this algorithm, which consists in proving that the algorithm always
terminates and that when it terminates it gives an mgu implies the
existence theorem. Several variants of this first-order unification
algorithm appear in well-known textbooks on computational and
mathematical logic, semantics of programming languages, rewriting
theory, etc. (e.g.,
\cite{Llo87,EbFlTh84,Bu98,Te2003,BaNi98,Hi97}). Since the presented
formalization follows the classical proof schema, no analytic
presentation of this proof is given here.

Basic notions on unification are specified straightforwardly in the
language of PVS.  For instance the notion of most general substitution
is given as

{\small
\begin{verbatim}
<=(theta, sigma): bool = EXISTS tau: sigma = comp(tau, theta)
\end{verbatim}
}

From this specification, one proves that the relation {\tt <=} is a
pre-order (i.e., reflexivity and transitivity).

The notions of unifier, unifiable, the set of unifiers of two terms
and a most general unifier of two terms are specified as

{\small
\begin{verbatim}
unifier(sigma)(s,t): bool = ext(sigma)(s) = ext(sigma)(t)
\end{verbatim}
} {\small
\begin{verbatim}
unifiable(s,t): bool = EXISTS sigma: unifier(sigma)(s,t)
\end{verbatim}
} {\small
\begin{verbatim}
U(s,t): set[Sub] = {sigma: Sub | unifier(sigma)(s,t)}
\end{verbatim}
} {\small
\begin{verbatim}
mgu(theta)(s,t): bool = 
  member(theta, U(s,t)) & FORALL sigma: member(sigma, U(s,t)) => theta <= sigma
\end{verbatim}}

  Several auxiliary lemmas related with the previous notions were also
  formalized as the ones presented below: {\tt unifier\_o} formalizes
  the fact that, whenever $\sigma\in U(s\hat{\theta},t\hat{\theta})$,
  $\sigma\circ\theta \in U(s,t)$; {\tt mgu\_o}, that whenever $\rho
  \geq \sigma$, $\rho\circ\theta \geq \sigma\circ\theta$; {\tt
    unifier\_and\_sub}, that instantiations of unifiers are unifiers;
  {\tt idemp\_mgu\_iff\_all\_unifier} that the idempotence property of
  mgu's holds, and; {\tt unifiable\_terms\_} {\tt unifiable\_args} formalizes the fact that corresponding
  subterms of unifiable terms are unifiable, .

  {\small
\begin{verbatim}
unifier_o: LEMMA
 member(sig, U(ext(theta)(s),ext(theta)(t))) => member(comp(sig,theta), U(s,t))
\end{verbatim}
  } {\small
\begin{verbatim}
mgu_o: LEMMA sig <= rho => comp(sig, theta) <= comp(rho, theta)
\end{verbatim}
  } {\small
\begin{verbatim}
unifier_and_subs: LEMMA
 member(theta, U(s,t)) => (FORALL (sig: Sub): member(comp(sig, theta), U(s,t)))
\end{verbatim}
  } {\small
\begin{verbatim}
idemp_mgu_iff_all_unifier: LEMMA FORALL (theta: Sub | member(theta, U(s,t))):
    mgu(theta)(s,t) & idempotent_sub?(theta)  <=>
    (FORALL (sig: Sub | member(sig, U(s,t))): sig = comp(sig, theta))
\end{verbatim}
  } {\small
\begin{verbatim}
unifiable_terms_unifiable_args: LEMMA
FORALL (s: term, t: term, p: position | positionsOF(s)(p) & positionsOF(t)(p)):
        member(sig, U(s,t)) => member(sig, U(subtermOF(s, p), subtermOF(t, p)))
\end{verbatim}
  }

  The unification algorithm receives two unifiable terms as arguments
  and is specified as the function {\tt unification\_algorithm},
  presented below. This function together with the two auxiliary
  functions {\tt sub\_of\_frst\_diff} and {\tt resolving\_diff}, to be
  explained in the remaining of this section, conform the kernel of
  the unification specified mechanism.

  {\small
\begin{verbatim}
unification_algorithm(s: term, (t: term | unifiable(s,t))): RECURSIVE Sub = 
IF s = t THEN identity ELSE LET sig = sub_of_frst_diff(s, t) IN
        comp( unification_algorithm((ext(sig))(s), (ext(sig)(t))), sig) ENDIF
MEASURE Card(union(Vars(s), Vars(t)))
\end{verbatim}
  }

  In this specification, the function {\tt sub\_of\_frst\_diff(s, t)},
  presented below, gives as result a substitution that resolves the
  first difference (left-most, outer-most in the structure of the
  terms) between the terms {\tt s} and {\tt t}, that are unifiable and
  different terms. In order to generate this substitution, the
  subterms that generate the difference must occur in the same
  position of {\tt s} and {\tt t}, one of these terms must be a
  variable and the other, a term without occurrences of this
  variable. The {\tt unification\_algorithm} recursive function has a
  pair of unifiable terms as domain type, given by the parameters {\tt
    s} and {\tt t}, and in the interesting case, after encountering
  the resolving substitution $\sigma$ for the first difference, it
  returns the composition of the result of the recursive call with the
  arguments {\tt s$\hat{\sigma}$} and {\tt t$\hat{\sigma}$} and
  $\sigma$.

  The functions {\tt resolving\_diff} and {\tt sub\_of\_frst\_diff},
  presented below, have the same type of parameters, and the former
  returns the first (left-most, outer-most) position of conflict
  between the unifiable and different terms {\tt s} and {\tt t}, as
  previously explained, while the latter returns the substitution that
  solves the conflict at the position generated by the function {\tt
    resolving\_diff}.

  {\small
\begin{verbatim}
resolving_diff(s: term, (t: term | unifiable(s,t) & s /= t ) ): RECURSIVE position =
 (CASES s OF
     vars(s): empty_seq,  
  app(f, st): IF length(st) = 0 THEN empty_seq
              ELSE (CASES t OF  
               vars(t): empty_seq,
          app(fp, stp): LET k: below[length(stp)] =
                min({kk: below[length(stp)] |
                    subtermOF(s,#(kk+1)) /= subtermOF(t,#(kk+1))}) IN
           add_first(k+1, resolving_diff(subtermOF(s,#(k+1)), subtermOF(t,#(k+1))))
                         ENDCASES)   ENDIF ENDCASES)
MEASURE s BY <<
\end{verbatim}
  }

  {\small
\begin{verbatim}
sub_of_frst_diff(s: term , (t: term | unifiable(s,t) & s /= t )): Sub =
  LET k: position = resolving_diff(s,t) IN
     LET sp = subtermOF(s,k) , tp = subtermOF(t,k) IN
        IF vars?(sp) THEN (LAMBDA (x: (V)): IF x = sp THEN tp ELSE x ENDIF)
        ELSE (LAMBDA (x: (V)): IF x = tp THEN sp ELSE x ENDIF) ENDIF 
\end{verbatim}
  }

  \subsection{Termination}\label{ssec:termination}

  Notice that the measure of the function {\tt unification\_algorithm}
  is the cardinality of the union of the sets of variables occurring
  in the term parameters {\tt s} and {\tt t}. From this measure, the
  PVS type-checker generates an interesting type proof obligation
  concerning the property of decreasingness of this measure, that
  guarantees the termination of the algorithm for all pairs of
  unifiable terms.

  {\small
\begin{verbatim}
unification_algorithm_TCC6: OBLIGATION FORALL (s, (t | unifiable(s, t))): 
NOT s = t =>  (FORALL (sig: Sub): sig = sub_of_frst_diff(s, t) =>
    Card(union(Vars(ext(sig)(s)), Vars(ext(sig)(t)))) < Card(union(Vars(s), Vars(t))))
\end{verbatim}
  }

  Although this key TCC is automatically generated, it is not
  automatically proved by PVS. In order to prove this TCC, one should
  first prove the following auxiliary lemma:

  {\small
\begin{verbatim}
vars_ext_sub_of_frst_diff_decrease: LEMMA
 FORALL (s: term, t: term | unifiable(s, t) & s /= t):
  LET sig = sub_of_frst_diff(s, t) IN
   Card(union( Vars(ext(sig)(s)), Vars(ext(sig)(t)))) < Card(union( Vars(s), Vars(t)))
\end{verbatim}
  }

  To prove the previous lemma, one requires the following additional
  lemma:

  {\small
\begin{verbatim}
union_vars_ext_sub_of_frst_diff : LEMMA
 FORALL (s : term, t : term | unifiable(s, t) & s /= t) :
  LET sig = sub_of_frst_diff(s, t) IN union(Vars(ext(sig)(s)), Vars(ext(sig)(t))) = 
                                      difference(union( Vars(s), Vars(t)), Dom(sig))
\end{verbatim}
  }

  The proof of the previous lemma requires that the substitution
  $\sigma$, that resolves the first conflict between the given terms,
  maps a variable into a term without occurrences of this variable.
  From this fact, it is possible to guarantee that the mapped variable
  disappears from the instantiated terms {\tt s$\hat{\sigma}$} and
  {\tt t$\hat{\sigma}$}, and hence the decreasing property holds. This
  is formalized as the lemma:

  {\small
\begin{verbatim}
 sub_of_frst_diff_remove_x : LEMMA FORALL (s:term, t:term | unifiable(s, t) & s /= t):
   LET sig = sub_of_frst_diff(s, t) IN Dom(sig)(x) =>  
         (NOT member(x, Vars(ext(sig)(s)))) AND (NOT member(x, Vars(ext(sig)(t))))
\end{verbatim}
  }

  Two other lemmas, one for $s$ and the other for $t$, formalize the
  fact that the variables in the $\sigma$ instantiated terms are
  contained in the set of variables occurring in the original terms
  being unified.

  {\small
\begin{verbatim}
  vars_sub_of_frst_diff_s_is_subset_union : LEMMA
    FORALL (s : term, t : term | unifiable(s, t) & s /= t):
        LET sig = sub_of_frst_diff(s, t) IN
           subset?(Vars(ext(sig)(s)), union( Vars(s), Vars(t)))
\end{verbatim}
  }

  Applying the previous lemmas, it is formalized the fact that the
  cardinality of the set of variables occurring in the terms being
  unified decreases after resolving each conflict between the terms.

  In the remaining of this section the formalization of lemma {\tt
    union\_vars\_ext\_sub\_of\_frst\_diff}, the lemma presented above, will be
  explained.  After a first step of skolemization and simplifications,
  the following sequent is obtained.

  {\small
\begin{verbatim}
{-1}  sub_of_frst_diff(s, t) = sig
  |-------
{1}   union(Vars(ext(sig)(s)), Vars(ext(sig)(t)))(x) IFF
       difference(union(Vars(s), Vars(t)), Dom(sig))(x)
\end{verbatim}
  }

  Note that there is a variable {\tt x}, resulting from an application
  of the PVS proof command ``decompose-equality'' that simplifies the
  equality between sets in the consequent formula into a
  biconditional, where the following assertion is established: $x$ is
  a member of $Vars(s\hat{\sigma}) \cup Vars(t\hat{\sigma})$ if, and
  only if, $x$ is a member of $Vars(s) \cup Vars(t) \setminus
  Dom(\sigma)$. At this point, a propositional simplification is
  applied and the proof is divided in two branches, presented below,
  one for each direction of the biconditional:

  \begin{itemize}

  \item $x \in Vars(s\hat{\sigma}) \cup Vars(t\hat{\sigma})$ implies
    $x \in Vars(s) \cup Vars(t) \setminus Dom(\sigma)$.

    After expanding the definitions of {\tt difference} and {\tt
      union}, the following sequent is obtained:

    {\small
\begin{verbatim}
{-1}  Vars(ext(sig)(s))(x) OR Vars(ext(sig)(t))(x)
[-2]  sub_of_frst_diff(s, t) = sig
  |-------
{1}   (Vars(s)(x) OR Vars(t)(x)) AND NOT Dom(sig)(x)
\end{verbatim}
    }

    Then, after propositional simplification, the proof divides into
    four branches:

    \begin{enumerate}

    \item In this case, $x \in Vars(s\hat{\sigma})$ and one should
      verify that either $x \in Vars(s)$ or $x \in Vars(t)$, which is
      done by application of lemma {\tt
        vars\_\-sub\_\-of\_\-frst\_\-diff\_s\_\-is\_\-subset\_\-union}.

    \item In this case, $x \in Vars(s\hat{\sigma})$ and one should
      verify that $x \notin Dom(\sigma)$, which is done by application
      of lemma {\tt sub\_of\_frst\_diff\_remove\_x}.

    \item[3, 4.] These cases are similar to the previous two cases for
      the term $t$.

    \end{enumerate}

  \item $x \in Vars(s) \cup Vars(t) \setminus Dom(\sigma)$ implies $x
    \in Vars(s\hat{\sigma}) \cup Vars(t\hat{\sigma})$.

    In this branch, after propositional simplification, one should
    verify that $x \in Vars(s)$ implies $x \in Vars(s\hat{\sigma})$
    or, $x \in Vars(t)$ implies $x \in Vars(t\hat{\sigma})$. This is
    true because if $x \notin Dom(\sigma)$, then for a position $\pi
    \in Pos(s)$ such that $s|_{\pi} = x$, one has
    $(s|_{\pi})\hat{\sigma} = (s\hat{\sigma})|_{\pi} = x$.

  \end{itemize}

\subsection{Soundness}
After establishing the termination of the specified function {\tt
  unification\_algorithm}, its correctness is formalized and applied
in order to prove the Theorem \ref{existence-mgu} of existence of
mgu's that is specified as:

{\small
\begin{verbatim}
unification: LEMMA unifiable(s,t) => EXISTS theta: mgu(theta)(s,t)
\end{verbatim}
}

This lemma is proved applying the two lemmas below.  The first one
states that the substitution given by the function {\tt
  unification\_algorithm} is, in fact, a unifier and the second one
that it is an mgu.

{\small
\begin{verbatim}
unification_algorithm_gives_unifier: LEMMA
    unifiable(s,t) => member(unification_algorithm(s, t), U(s, t))
\end{verbatim}
} {\small
\begin{verbatim}
unification_algorithm_gives_mg_subs: LEMMA
    member(rho, U(s, t)) => unification_algorithm(s, t) <= rho
\end{verbatim}
}

The formalization of the lemma {\tt
  unification\_algorithm\_gives\_unifier} is done by induction on the
cardinality of the set of variables occurring in {\tt s} and {\tt
  t}. For proving this lemma three auxiliary lemmas are necessary:

\begin{itemize}
\item the lemma {\tt vars\_ext\_sub\_of\_frst\_diff\_decrease}
  described in the previous subsection, which guarantees that the set
  of variables decreases;
\item {\small
\begin{verbatim}
ext_sub_of_frst_diff_unifiable: LEMMA
    FORALL (s: term, t: term | unifiable(s, t) & s /= t):
        LET sig = sub_of_frst_diff(s, t) IN unifiable(ext(sig)(s), (ext(sig)(t)))
\end{verbatim}
  } which states that the instantiations of two different and
  unifiable terms {\tt s}$\hat{\sigma}$ and {\tt t}$\hat{\sigma}$ with
  the substitution $\sigma$ that resolves the first conflict between
  these terms, are still unifiable; and

\item the lemma {\tt unifier\_o}, presented at the beginning of this
  section, which states that for any unifier $\theta$ of {\tt
    s}$\hat{\sigma}$ and {\tt t}$\hat{\sigma}$, $\theta\circ\sigma$ is
  a unifier of {\tt s} and {\tt t}.
\end{itemize}

The formalization of the lemma {\tt
  unification\_algorithm\_gives\_mg\_subs} is done by induction on the
cardinality of the set of variables occurring in {\tt s} and {\tt t}
too. For proving this lemma two auxiliary lemmas are applied: the
lemma {\tt vars\_ext\_sub\_of\_frst\_diff\_decrease} and the lemma
presented below, which states that for each unifier $\rho$ of {\tt s}
and {\tt t}, two different and unifiable terms, and given $\sigma$ the
substitution that resolves the first difference between these terms,
there exist $\theta$ such that $\theta\circ\sigma = \rho$.
                
{\small
\begin{verbatim}
sub_of_frst_diff_unifier_o: LEMMA FORALL (s:term, t:term | unifiable(s, t) & s /= t):
 member(rho, U(s, t)) =>
           LET sig = sub_of_frst_diff(s, t) IN EXISTS theta: rho = comp(theta, sig)
\end{verbatim}
}

In the remaining of this section the formalization of {\tt
  sub\_of\_frst\_diff\_unifier\_o} will be explained.

It should be proved that $\theta \circ \sigma$ and $\rho$ map each
variable {\tt x} in their domain, that should be the same set of
variables, into the same terms.  The formalization starts by a
skolemization and then, in order to provide a name, {\tt p}, for the
position in which the first difference between terms {\tt s} and {\tt
  t} is detected, an application of the PVS proof command ``name'' is
done. In this way the additional premise {\tt resolving\_diff(s, t) =
  p} is included.

{\small
\begin{verbatim}
   {-1}  resolving_diff(s, t) = p   [-2]  sub_of_frst_diff(s, t) = sig
   [-3]  member(rho, U(s, t))
     |-------
   [1]   EXISTS theta: rho = comp(theta, sig)
\end{verbatim}
}

The proof strategy is to instantiate the existential formula in the
consequent with $\rho$ itself, having in mind that if $\rho \in U(s,
t)$ then $\rho \in U(s|_q, t|_q)$, for any valid position $q$ of $s$
and $t$, and in particular, for the position of the first detected
difference $p$. It is known that at position $p$, either $s|_p$ or
$t|_p$ should be a variable; so the strategy is to analyze both
possible cases. The sequent below is obtained in the case in which
$s|_p$ is a variable. In this sequent {\tt x} is an arbitrary
variable.

{\small
\begin{verbatim}
   {-1}  vars?(subtermOF(s, p))       [-2]  ext(rho)(subtermOF(s, p)) = 
         ext(rho)(subtermOF(t, p))    [-3]  resolving_diff(s, t) = p   
   [-4]  sub_of_frst_diff(s, t) = sig [-5]  ext(rho)(s) = ext(rho)(t)
     |-------
   [1]   rho(x) = ext(rho)(sig(x))
\end{verbatim}
}

The variable {\tt x} in the consequent of this sequent appears after
an application of the PVS proof command ``decompose-equality'' that
simplifies equality between substitutions into equality of the
application of the substitutions to any variable: $\rho \circ \sigma =
\rho$, whenever for any $x$ $(x\sigma)\hat{\rho} = x\rho$.

The proof is obtained by case analysis: when $x = s|_p$ and when $x
\neq s|_p$,

\begin{itemize}
\item In the former case, the formula {\tt x = subtermOF(s, p)} is
  added to the antecedents.

  Note that $(s|_p)\hat{\sigma} = t|_p$, that is {\tt
    ext(sig)(subtermOF(s, p)) = subtermOF(t, p)}, by definition of
  {\tt sub\_of\_frst\_diff}, and $(s|_p)\hat{\rho} =
  (t|_p)\hat{\rho}$. Then, one can conclude that $(s|_p)\hat{\rho} =
  ((s|_p)\hat{\sigma})\hat{\rho}$. But, in this case, $x = s|_p$;
  thus, one can complete this branch of the proof expanding the
  definition of {\tt sub\_of\_frst\_diff} with an application of the
  proof command ``expand'' and making simplifications with the
  commands ``replace'' and ``assert''.

\item In the latter case, the sequent to be considered is presented
  below.  Notice that the negated equality that characterizes this
  case is positively presented as a consequent of the sequent.
             
  {\small
\begin{verbatim}
   [-1]  vars?(subtermOF(s, p))         [-2]  ext(rho)(subtermOF(s, p)) = 
         ext(rho)(subtermOF(t, p))      [-3]  resolving_diff(s, t) = p
   [-4]  sub_of_frst_diff(s, t) = sig   [-5]  ext(rho)(s) = ext(rho)(t)
     |-------
   {1}   x = subtermOF(s, p)            [2]   rho(x) = ext(rho)(sig(x))
\end{verbatim}
  }

  In this case, note that $x$ does not belong to the domain of
  substitution $\sigma$, because the domain of $\sigma$ is the
  singleton $\{s|_p\}$. Then $x\sigma = x$. Therefore the equality
  $x\rho = (x\sigma)\hat{\rho}$ is true, which is sufficient to
  complete this branch of the proof.
\end{itemize}

At this point, two cases remain to be considered: the case where
$t|_p$ is a variable, that is formalized in a way entirely analogous
to the previous case, and the case where neither $s|_p$ nor $t|_p$ are
variables.

In the latter case, again one should apply that at a conflicting
position of two unifiable terms it is impossible that none of the
subterms is a variable. This result was already formalized as a lemma
called {\tt resolving\_diff\_vars}. Then using this lemma and
instantiating appropriately one obtains the sequent:

{\small
\begin{verbatim}
 {-1}  p = resolving_diff(s, t) => vars?(subtermOF(s, p)) OR vars?(subtermOF(t, p))
 [-2]  resolving_diff(s, t) = p
   |-------
 [1]   vars?(subtermOF(t, p))   [2]   vars?(subtermOF(s, p))
\end{verbatim}
}

In this sequent the contradiction is already established, and can be
captured with a simple application of the PVS proof command
``assert''. The proof of the main lemma used in this branch of the
proof, {\tt resolving\_diff\_vars}, previously mentioned, follows by
induction on the structure of the term $s$ as explained below.

If $s$ is a variable, the position $p$ should be the root position
{\tt empty\_seq} and at this position, the term $s|_\varepsilon$ is a
variable.  If $s$ is an application, the proof follows by expanding
the definition of {\tt resolving\_diff} and considering the three
possible cases, namely:

\begin{enumerate}

\item $s$ is a constant. Then the position of the first difference
  should be $\varepsilon$ and $t$ should be a variable, since the
  terms are unifiable.

\item $s$ is a non constant application and $t$ is a variable. Similar
  to the previous case.

\item $s$ is a non constant application and $t$ is an application.
  The position $p$ cannot be the root position. The sequent
  corresponding to this case is presented below.

  {\small
\begin{verbatim}
{-1}  p = add_first(k, resolving_diff(subtermOF(s,  #(k)), subtermOF(t,  #(k))))
[-2]  FORALL (x: below[args(s)`length]):
       FORALL (t: term | unifiable(args(s)`seq(x), t) & args(s)`seq(x) /= t,
               p: position | positionsOF(args(s)`seq(x))(p) & positionsOF(t)(p)):
          p = resolving_diff(args(s)`seq(x), t) =>
           vars?(subtermOF(args(s)`seq(x), p)) OR vars?(subtermOF(t, p))
  |-------
[1]   vars?(t)                 [2]   length(args(s)) = 0
[3]   vars?(subtermOF(s, p))   [4]   vars?(subtermOF(t, p))
\end{verbatim}
  }

  In this sequent the induction hypotesis, that is the antecedent
  formula [-2], should be instantiated with {\tt k - 1}, in order to
  capture the subterm of $s|_k$, i.e., the $(k - 1)$-th element of the
  sequence of arguments of the root symbol of $s$, {\tt args(s)`seq(k
    - 1)}. Then, the position $p$ equals the concatenation of $k$ with
  the first difference between terms $s|_k$ and $t|_k$, here denoted
  as $p = k \circ q$. By induction hypotesis either $(s|_k)|_q$ or
  $(t|_k)|_q$ is a variable. But $(s|_k)|_q = s|_{k \circ q}$ and
  $(t|_k)|_q = t|_{k \circ q}$, which concludes the proof.

\end{enumerate}

\subsection{Verification of unification algorithms}

This methodology of proof of the existence of mgu's can be applied in
order to formalize the completeness of unification algorithms {\em \`a
  la} Robinson, as presented in detail in \cite{AMARG2010} for a
greedy unification algorithm.  This is illustrated in the
\emph{theory} {\tt \bf robinsonunification} also available inside {\tt
  trs} as well as in a more recent efficient specification {\tt\bf
  robinsonunificationEF} (see the {\tt trs} hierarchy in
Fig. \ref{figUnificationInsideTRS}).

The main functions in the \emph{theory} {\tt \bf robinsonunification}
are: {\tt first\_diff}, {\tt link\_of\_frst\_diff} and {\tt
  robinson\_unification\_algorithm} whose roles are analogous
respectively to the ones of the functions {\tt resolving\_diff}, {\tt
  sub\_of\_frst\_diff} and {\tt unification\_algorithm}.  These
functions are specified in such a way that whenever unsolvable
differences are detected (by the function {\tt first\_diff}) the
substitution ``{\tt fail}'' is returned. This substitution is built
explicitly as the substitution with the singleton domain {\tt \{xx\}}
and image {\tt ff(xx)}, where {\tt xx} and {\tt ff} are, respectively,
a constant and a unary function. In this way, the substitution {\tt
  fail} is discriminated from any other possible unifier which is
built by the function {\tt robinson\_unification\_algorithm} for all
pair of terms.

The function {\tt link\_of\_frst\_diff}, presented below, either
builds the resolving link substitution for the first difference whose
position is detected by {\tt first\_diff} or returns {\tt
  fail}. According to these two options, the function {\tt
  robinson\_unification\_algorithm}, also presented below, either
builds the mgu or returns {\tt fail}.

{\small
\begin{verbatim}
link_of_frst_diff(s : term , (t : term | s /= t )) : Sub =
  LET k : position = first_diff(s,t) IN
     LET sp = subtermOF(s,k) , tp = subtermOF(t,k) IN 
        IF vars?(sp) 
        THEN IF NOT member(sp, Vars(tp))
             THEN (LAMBDA (x : (V)) : IF x = sp THEN tp ELSE x ENDIF)
             ELSE fail ENDIF 
        ELSE  IF vars?(tp) THEN 
                IF NOT member(tp, Vars(sp))
                THEN (LAMBDA (x : (V)) : IF x = tp THEN sp ELSE x ENDIF)
                ELSE fail ENDIF           
              ELSE fail ENDIF  ENDIF

robinson_unification_algorithm(s, t : term) : RECURSIVE Sub =
 IF s = t THEN identity  
 ELSE LET sig = link_of_frst_diff(s,t) IN
  IF sig = fail THEN fail
  ELSE LET sigma = robinson_unification_algorithm(ext(sig)(s) , ext(sig)(t)) IN
   IF sigma = fail THEN fail ELSE comp(sigma, sig) ENDIF
  ENDIF  ENDIF
 MEASURE Card(union(Vars(s), Vars(t)))
\end{verbatim}
}

The \emph{theory} {\tt \bf robinsonunification} consists of 47 lemmas
from which 24 are TCCs.  The specification file has 249 lines and its
size is 8.6 KB, and the whole proof file has 12091 lines and 739 KB
and was described in detail in \cite{AMARG2010}.

The \emph{subtheory} {\tt \bf robinsonunificationEF} includes an
``efficient'' version of the unification algorithm in which after
resolving each conflicting position between two terms the next
conflict is searched starting from the position of conflict previously
resolved instead from the root position of the instantiated terms as
it is done in the \emph{theories} {\tt \bf unification} and {\tt \bf
  robinsonunification}.  The main functions found in this improved
version of the algorithm are {\tt next\_position} and {\tt
  robinson\_unification\_algorithm\_aux}.

The function {\tt next\_position} takes as arguments two terms and a
valid position $\pi$ of both terms, and returns the next conflicting
position. Once all differences between the terms occurring in previous
positions to $\pi$ (left-most, outer-most) and at position $\pi$
itself have been resolved, the next conflict should occur in a
position to the right, and therefore there is no need to scan again
the instantiated terms starting from the root position.

The function {\tt robinson\_unification\_algorithm\_aux} also takes as
arguments two terms and a position of these terms, and returns a
substitution, but now in the process of unification the next conflict
position is fetched from the first position of conflict, using the
function {\tt next\_position}.

{\small
\begin{verbatim}
 next_position(s, t : term, 
               p : position | positionsOF(s)(p) AND positionsOF(t)(p)): 
 RECURSIVE position = 
 IF p = empty_seq THEN empty_seq 
 ELSE LET pi0 = delete(p,length(p) - 1) IN
      IF f(subtermOF(s,pi0)) /= f(subtermOF(t,pi0)) THEN pi0 
      ELSE LET pi = add_last(delete(p, length(p) - 1), last(p) + 1) IN
           IF positionsOF(s)(pi) THEN 
            IF subtermOF(s, pi) /= subtermOF(t, pi) THEN pi 
            ELSE next_position(s,t, pi) ENDIF
           ELSE IF pi0 /= empty_seq THEN next_position(s, t, pi0)
                ELSE empty_seq ENDIF 
           ENDIF  ENDIF  ENDIF  
 MEASURE IF p = empty_seq THEN lex2(0,0)
         ELSE lex2(length(p),
                    arity(f(subtermOF(s, delete(p,length(p) - 1)))) - last(p))
         ENDIF

 robinson_unification_algorithm_aux(s, t : term, p : position | 
    positionsOF(s)(p) AND  positionsOF(t)(p)) : RECURSIVE Sub = 
 IF subtermOF(s,p) = subtermOF(t,p) THEN 
   LET pi = next_position(s, t, p) IN
      IF pi = empty_seq THEN identity
      ELSE robinson_unification_algorithm_aux(s,t,pi)
      ENDIF
 ELSE LET sig = link_of_frst_diff(subtermOF(s,p),subtermOF(t,p)) IN
      IF sig = fail THEN fail
      ELSE LET pi = next_position(ext(sig)(s), ext(sig)(t), 
                         p o first_diff(subtermOF(s,p),subtermOF(t,p))) IN
           IF pi = empty_seq THEN  sig
           ELSE LET sigma = robinson_unification_algorithm_aux(
                                        ext(sig)(s), ext(sig)(t), pi) IN
                IF sigma = fail THEN fail ELSE comp(sigma, sig) ENDIF
           ENDIF  ENDIF  ENDIF 
 MEASURE  lex2(Card(union(Vars(s), Vars(t))), Card(right_pos(s,p)))
\end{verbatim}}

  Formalization of correctness of this specification requires several
  additional effort and, in particular, specialized inductive proof
  that are based on the more elaborated measures necessaries for the
  previous two functions.

  \section{Related work}

  Correctness of unification algorithms has been the center of several
  formalizations in a variety of theorem provers. Starting from a
  formalization in LCF \cite{Pa1985}, other formal proofs have been
  given, for example, in Isabelle/HOL, Coq
  \cite{Rou1992,BlKo2011,KoCa2009}, ALF \cite{Bove1999} and ACL2
  \cite{RRMMAH2006}.

  The earlier LCF formalization of the unification algorithm was given
  by Paulson \cite{Pa1985}. Paulson's approach was followed by Konrad
  Slind in the theory {\tt Unify} formalized in Isabelle/HOL from
  which an improved version called {\tt Unification} is available
  now. Unlike other approaches, in Slind's formalization as in the
  presented here idempotence of the computed unifiers is unnecessary
  to prove neither termination nor correctness of the specified
  unification algorithm.  In contrast with our textbook style
  termination proof, which is based on the fact that the number of
  different variables occurring in the terms being unified decreases
  after each step of the unification algorithm (Section
  \ref{ssec:termination}), the termination proof of the theory {\tt
    Unify} is based on separated proofs of \emph{non-nested} and
  \emph{nested} termination conditions and the unification algorithm
  is specified taking as basis a specification of terms built by a binary
  combinator operator ({\tt Comb}).

  Recent Coq formalizations of unification algorithms were presented
  in \cite{BlKo2011} and \cite{KoCa2009}. The formalization in
  \cite{BlKo2011} is part of a library called CoLoR, and the most
  significant difference is that here substitutions are specified as
  finite maps from unrestricted variables into general terms, whereas
  in CoLoR they are specified as functions from type variables to a
  generalized term structure.  In \cite{KoCa2009}, Kothari and
  Caldwell presented a specification of a unification algorithm for
  equalities in the language of simple types. This kind of unification
  has direct applications in type inference algorithms. This
  unification algorithm is proved correct by showing that it satisfies
  four axioms: that the computed mgu is a unifier; that it is in fact
  a most general unifier; that its domain is restricted to the set of
  free variables in the input equational problem and that the theorem
  of existence of mgu's holds. In a later work, the same authors
  showed that three additional axioms, being one of them idempotence
  of mgu's, are also satisfied. Since simple types are built in a
  language of symbols for basic types and a unique binary operator
  symbol, ($\to$), the current approach can be directly applied to the
  restricted language of simple types treated in \cite{KoCa2009}.  An
  additional fact that makes the current formalization closer to the
  usual theory of unification as presented in well-known textbooks
  (e.g., \cite{Llo87, BaNi98}), is the decision to specify terms as a
  data type built from variables and the operator {\tt app} that
  builds terms as an application of a function symbol (of a given
  arity) to a sequence of terms with the right length. In this way,
  substitutions were specified as a function from variables to terms
  and the construction of the homomorphic extension results
  straightforward.

  Earlier related work in Coq includes~\cite{Rou1992}, where an
  algorithm similar to Robinson's one was extracted from a
  formalization that uses a generalized notion of terms, that uses
  binary constructors in the style of Manna and Waldinger, whose
  translation to the usual notation is not straightforward. More
  recently, in \cite{Constable09}, a certified resolution algorithm
  for the propositional calculus is extracted from a Coq specification
  that requires unification of propositional expressions.

  In \cite{Bove1999} a formalization of a first-order unification
  algorithm is given. The main difference with the current formalization is
  that here one defines the application of a substitution to a term only by
  recursion on the term, and there the author defines the application of a
  substitution to a term in two ways: by recursion on the term
  (parallel application) and by recursion on the substitution
  (sequential application). Thus, for a given substitution and a given
  term, the application of the substitution to the term might result
  in different terms, depending on whether one follows the definition
  of the parallel application or the sequential application. However,
  both applications give the same result for idempotent
  substitutions. In other words, unlike the current approach, idempotence of
  the computed unifiers is necessary to prove the correctness of the
  specified unification algorithm.

  In \cite{RRMMAH2006} a formalization of the correctness of an
  implementation of an $O(n^2)$ run-time unification algorithm in ACL2
  is presented. The specification is based on Corbin and Bidoit's
  development \cite{CoBi83} as presented in \cite{BaNi98} in which
  terms are represented as directed acyclic graphs (DAGs). The merit
  of this formalization is that by taking care of an specific data
  structure such as DAGs for representing terms, the correctness proof
  results much more elaborated than the current one. But in the
  current paper, the focus is to have a natural mechanical proof of
  the existence of mgu's, that is the strictly necessary in a
  formalization of the correctness of the Critical Pair Knuth-Bendix
  theorem. Although the representation of terms is sophisticated (via
  DAGs), the referred formalization diverges from textbooks proofs of
  correctness of the unification algorithm in which it is first-order
  restricted. In fact, instead representing second-order objects such
  as substitutions as functions from the domain of variables to the
  range of terms, they are specified as first-order association
  lists. In our approach, taking the decision to specify substitutions
  as functions allows us to apply all the theory of functions
  available in the higher-order proof assistant PVS, which makes our
  formalization very close to the ones available in textbooks.

  As mentioned in the introduction, as part of the PVS \emph{theory}
  {\color{blue} \tt trs} presented in \cite{GaAR2008b} there are
  formalizations of non-trivial results on rewriting, such as the
  well-known Knuth-Bendix Critical Pair Theorem, that
  requires the theorem of existence of mgu's. The style of formalization
  of existence of mgu's can be followed in order to verify the soundness
  and completeness of unification algorithms \`{a} la Robinson, as
  illustrated in \cite{AMARG2010} for a greedy algorithm.
  The proof methodology used to prove termination and soundness in the
  formalization of the theorem of existence of mgu's is adapted in
  order to verify the correctness of unification algorithms as
  described in \cite{AMARG2010}.

  \section{Conclusions and Future Work}\label{Sec:conclusion}

  A formalization developed in the language of the proof assistant PVS
  of the theorem of existence of mgu's was presented. This
  formalization is close to textbooks proofs and was applied to
  present a complete formalization of the well-know Knuth-Bendix
  Critical Pair theorem.  The methodology of proof can be directly
  applied in order to certify the correctness of first-order
  unification algorithms \`a la Robinson.

  As future work, it is of great interest the extraction of certified
  unification algorithms alone or in several contexts of its possible
  applications such as the ones of first order resolution and of type
  inference.

  \bibliographystyle{eptcs}

\end{document}